\documentstyle[prl,aps,epsfig,twocolumn]{revtex}

\begin{document}
\title{Quantum calculations of Coulomb reorientation for sub-barrier fusion}

\author{C. Simenel $^{a,b)}$, Ph. Chomaz $^{a)}$ and G. de France $^{a)}$}
\address{
a) GANIL, BP 55027, F-14076 Caen Cedex 5, France \\
b) DSM/DAPNIA CEA SACLAY, F-91191 Gif-sur-Yvette, France
}

\date{\today}

\wideabs{
\maketitle
\begin{abstract}
Classical mechanics and Time Dependent Hartree-Fock (TDHF) calculations of heavy
ions collisions are
 performed to study the rotation of a deformed nucleus in the Coulomb
field of its partner. This reorientation is shown to be {\it
independent }on charges and relative energy of the partners. It
only depends upon the deformations and inertias. TDHF calculations
predict an increase by 30\% of the induced rotation due to quantum
effects while the nuclear contribution seems negligible. This
reorientation modifies strongly the fusion cross-section around
the barrier for light deformed nuclei on heavy collision partners.
For such nuclei a hindrance of the sub-barrier fusion is
predicted.
\end{abstract}

\pacs{Valid PACS appear here}
}
\smallskip Tunneling, the slow ''quantum leak'' through a classical barrier,
is an intriguing phenomenon in nature. In 1928, Gamow discovered
this effect looking for an explanation of the $\alpha$
radioactivity \cite{gam28}. However, the tunneling of complex
systems remains to be understood. As in the Gamow times nuclear
physics is providing one of the most challenging field to
understand tunneling. In particular, fusion cross sections
involving massive nuclei around the Coulomb barrier can be orders
of magnitude over one dimensional quantum tunneling predictions.
Couplings between the internal degrees of freedom and the relative
motion deeply modifies tunneling \cite{das83}. Neutron transfer,
excitation of low-lying vibrational and rotational states, neck
formation, zero-point motion and polarization of collective
surface vibration as well as static deformation have been
identified as key inputs in the understanding of this sub-barrier
fusion enhancement \cite{das98}.

For nuclei with a significant static quadrupole deformation \cite
{rie70,jen70}, the main effects are i) on the barrier height
(geometrical effect) since it is lower in the elongated direction
and ii) on the reorientation of the deformed nucleus (rotational
effect) under the torque produced by the long-range Coulomb force.
In \cite {sto78,gre86,wei91,lei93,mor94}, fusion excitation
functions were measured for $^{16}$O (spherical) + $^{144-154}$Sm
reactions. $^{144}$Sm is spherical whereas $^{154}$Sm is prolate ($\beta _{2}\!%
\approx \!0.3$). The data near the barrier were interpreted as
arising from the different orientations of the prolate nucleus. An
enhancement of the fusion probability is observed when the
deformation axis is parallel to the collision axis (''parallel
collision'') and a hindrance when the two axis are nearly
perpendicular (''perpendicular collision''). In these studies,
however, the assumption of an {\it isotropic orientation
distribution} of the deformed nucleus at contact was made. This
contradicts classical calculations \cite{hol69,wil67} which show
partial reorientation. From the quantum mechanics point of view,
the reorientation is a consequence of the excitation of rotational
states, which may affect near barrier fusion specially for light
deformed nuclei \cite{das94,bab00}. Computational techniques have
been developed in the past to solve coupled channel { (CC)
}equations for Coulomb excitation \cite{rho80,tol79,ros74,raw76}
but a good understanding of the Coulomb reorientation dynamics
during the approach phase is still required.

In this work we obtain a deeper insight in the Coulomb
reorientation revisiting the classical result,  solving
analytically and numerically the equations of motion of a rigid
body, and performing novel quantum approaches describing the
deformed projectile within the time dependent Hartree-Fock (TDHF)
theory. Then we  use these reorientation results  in a calculation
of the fusion cross-sections. { The induced effects give an
helpful interpretation of full CC calculations.}

Assuming first a classical treatment of nuclear orientation (the
various orientations do not interfere) and ignoring any
reorientation effect, the fusion cross section is given by the
average orientation formula (AOF) \cite{rum01,nag86}
\begin{equation}
\sigma _{fus.}\approx \int_{\varphi =0}^{\frac{\pi }{2}}\!\!\!\sigma
(\varphi )\sin \varphi \,d\varphi \,  \label{av_formul}
\end{equation}
where $\varphi $ is the angle between the deformation and the collision axis
and $\sigma (\varphi )$ is the associated cross section. However, the
Coulomb force induces a torque which, integrated over the whole history, up
to the distance of closest approach $D_0$, rotates the initial angle $%
\varphi _{\infty }$ into $\varphi _{0}.$ Because of reorientation $\Delta
\varphi =\varphi _{0}-\varphi _{\infty }\neq 0$, the distribution of $%
\varphi _{0}$ looses its isotropy and the $\sin \varphi $ term in Eq. \ref
{av_formul} has to be modified.

To estimate $\Delta \varphi$, we first consider the classical motion of a
deformed rigid projectile in the Coulomb field of the target. We assume that
the projectile of mass ${A_{p}}$ presents a sharp surface at a radius $%
R(\theta)\!\!=\!\!R_{0}\sqrt{\alpha ^{-4}\cos ^{2}\theta +\alpha ^{2}\sin
^{2}{\theta }}$ where $R_{0}\!\!=\!r_{0}{A_{p}}^{\frac{1}{3}}$, $%
r_{0}\!=\!1.2$~fm, $\alpha\!\! =\!\!1\!-\!\varepsilon $ and $\varepsilon
\!\!=\!\!\sqrt{5/16\pi} \beta _{2}$, the deformation parameter.

\begin{figure}[tbp]
\begin{center}
    \epsfig{figure=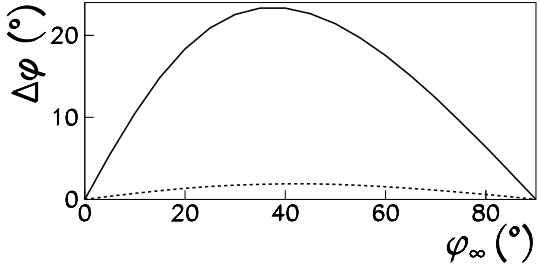,width=8.cm}
\end{center}
\caption{Classical calculations (started at $D_{\infty }=241$
fm) of the reorientation $\Delta \varphi $ of the projectile as
function of its initial orientation $\varphi _{\infty }$ for the
reactions $^{24}$Mg$+^{208}$Pb (solid line) and
$^{154}$Sm$+^{16}$O (dashed line) at the barrier.}
\label{delta_phi}
\end{figure}

Fig.~\ref{delta_phi} shows the reorientation as function of
the initial orientation for central collisions at the barrier $%
B\!\!=\!\!Z_{p}Z_{t}e^{2}/r_{0}\left(\!
A_{t}^{1/3}\!\!+\!A_{p}^{1/3}\!\right)$ where $Z_{p}$ ($A_{p}$) and $Z_{t}$ (%
$A_{t}$) are the projectile and target number of protons (nucleons). The
figure presents two typical asymmetric reactions of a prolate projectile on
a spherical target: $^{24}$Mg($\beta _{2}\!\!\approx\! 0.4$)$+^{208}$Pb and $%
^{154}$Sm($\beta _{2}\!\!\approx\!0.3$)$+^{16}$O. For symmetry reasons $%
\Delta \varphi \!=\!0{{}^{\circ }}$ for $\varphi_{\infty}\!
=\!0{{}^{\circ }} $ and $90{{}^{\circ}}$. The maximal
reorientation $\Delta \varphi _{max}$
occurs around $45{{}^{\circ }}.$ For the heavy deformed projectile, $^{154}$%
Sm, $\Delta \varphi _{max}$ is less than $2{{}^{\circ }}$ whereas for $^{24}$%
Mg it is large ($\sim\! 23{{}^{\circ }}$).

To understand this difference, an analytical expression for the
reorientation can be derived following the approximations { of}
ref. \cite{alder} i.e.  assuming that $\varepsilon $ and $\Delta
\varphi$ are small. Computing the torque leads to following the
equation of motion for $\varphi (t)$
\begin{equation}
\ddot{\varphi}(t)\approx \frac{9Z_{p}Z_{t}e^{2}\varepsilon }{2mA_{p}D(t)^{3}}%
\sin(2\varphi(t))  \label{eq:phi_second}
\end{equation}
where $m$ is the nucleon mass and $D(t)$ the distance between the two
nuclei. Replacing the time variable by $\xi (t)=D(t)/D_{0}$ and neglecting
deformation and rotation on the dynamics of $D(t)$, Eq. \ref{eq:phi_second}
becomes
\begin{equation}
\frac{\partial \varphi (\xi )}{\partial \xi }+2\xi \left( \xi -1\right)
\frac{\partial ^{2}\varphi (\xi )}{\partial \xi ^{2}}=\frac{9\varepsilon } {%
2\xi }\frac{A_{t}}{A_{p}+A_{t}}\sin(2\varphi_\infty)  \label{eq_diff}
\end{equation}
where $\sin(2\varphi(t))$ have been replaced by $\sin(2\varphi_\infty)$
treating the reorientation perturbatively (see \cite{alder}).  Only the
factor $A_{t}/(A_{p}+A_{t})$ remains since the initial center of mass energy
$E$ and the charges have been taken into account in $D_{0}=e^{2}Z_{p}Z_{t}/E$%
. The solution of Eq. \ref{eq_diff} is
\begin{equation}
\varphi (\xi )=\varphi_{{\small \infty }}\! +\! \frac{3 \varepsilon A_{t}}{%
A_{p}+A_{t}} \! \sin\!(2\varphi_{{\small \infty}})\!\left( \xi
\left(2\!-\!\zeta \right)\!-\!\ln\!\zeta\! +\!\delta \right)
\label{orientation}
\end{equation}
where $\zeta=1+\sqrt{1-\xi^{-1} }$ and $\delta =\ln 2-1/2$. Solved up to the
distance of closest approach ($\xi =1$), it leads to
\begin{equation}
\Delta \varphi =3 \varepsilon \frac{A_{t}}{A_{p}+A_{t}}\sin(2\varphi_\infty)%
\left( \frac{1}{2}+\ln 2\right) .  \label{reorientation}
\end{equation}

It can be shown by performing the time integral introduced in ref.
\cite {broglia} that Eq. \ref{reorientation} is equivalent to the
one reported in ref. \cite{broglia}. { It should be noticed that
Eq. \ref{reorientation} can  be transformed in order to explicitly
introduce the projectile moment of inertia $\mathcal I$ or
equivalently the rotational excitation energy
 of the first excited state ${ E}_{2}=6\hbar^2/2\mathcal I$.
 This clearly shows that in addition
to a non zero deformation, a finite $\mathcal I$ (i.e. a non-zero
${ E}_{2}$) is needed to get a reorientation, i.e. a deviation
from the AOF.}

Eq. \ref{reorientation} shows that the reorientation depends
neither on projectile and target charges nor on the relative
energy but only on the deformation and the mass ratio{  which is
nothing but an inertia factor}. This counter-intuitive result,
which has been exhaustively checked numerically, can be
understood: the increase of the Coulomb interaction with charges
(like $Z_{p}Z_{t}$) is compensated by an increase of $D_{0}$. A
similar balance occurs with incident energy. An increase in $E$
reduces $D_{0}$ and the time to interact leading to a zero net
effect on the integrated reorientation. The strong difference
between the two systems
shown in Fig.~\ref{delta_phi} is thus only due to the difference in $%
A_{t}/(A_{p}+A_{t})$, and not to the Coulomb forces.

\begin{figure}[tbp]
\begin{center}
    \epsfig{figure=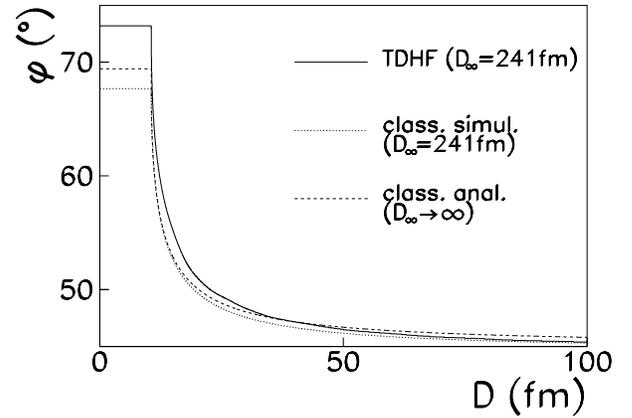,width=8.cm}
\end{center}
\caption{For the central collision of $^{24}$Mg on $^{208}Pb$ at
the barrier the orientation $\varphi $ of the $^{24}$Mg as
function of the relative distance $D$ predicted by TDHF with the
Sk$M^*$ force (solid line) or by classical simulation (dotted
line) and its analytic approximation (dashed line).} \label{phi_r}
\end{figure}

Fig.~\ref{phi_r} shows the evolution of { $\varphi(D)$ }
for the central reaction $^{24}$Mg$+^{208}$Pb at the barrier { with $%
\varphi_\infty =45{{}^{\circ }}$}. Results from the numerical
solution of the classical dynamic of a deformed rigid body (dotted
line) and approximated analytical expression Eq. \ref{orientation}
(dashed line) are very close. The small difference observed at the
turning point can be attributed to the higher orders terms in
$\varepsilon$. The difference at large distance is due to the {
finiteness of } $D_{\infty }$ (here $D_{\infty }=241$ fm) { in the
numerical simulation} while the analytical result integrates the
effects from $D_{\infty }\rightarrow \infty $.

To take into account the quantal nature of the nuclei and to avoid
the rigid body approximation, we have performed TDHF calculations
\cite {har28,foc30,vau72,bon76,neg82} of this nuclear reaction.
TDHF is optimized for the prediction of the average values of one
body observables { like deformation and orientation}. The
evolution of the one-body density matrix $\rho
=\sum_{n=1}^{N}\left| \varphi _{n}\right\rangle \left\langle
\varphi _{n}\right| $ is determined by a
Liouville equation, $i\hbar \partial _{t}\rho =[h(\rho ),\rho ]$ where $%
h(\rho )$ is the mean-field Hamiltonian. We { use} the code built
by P. Bonche and coworkers \cite{kim97} with an effective Skyrme
mean-field \cite{sky56}. { This code, which does not include
pairing, computes the evolution of a Slater determinant in a 3
dimensional box. The step size of the network is 0.8 fm and the
step time 0.45 fm/c.} Two different parametrizations were used,
Sk$M^{*}$ \cite{bar82} and SLy4 \cite{Cha98}, in order to control
that the conclusions are almost independent on the force. Because
of the long range nature of the Coulomb interaction, the
calculation must be started much before the turning point
typically for $D_\infty$ around $200$ fm. However, for reactions
below the barrier we can separate the dynamics of the target from
the one of the projectile. Therefore, we have modified the TDHF
code in order to compute the evolution of the nuclei separately in
their center of mass frame. The chosen box for $^{24}$Mg is a cube
of side size 16 fm. We assume that the centers of mass follow
Rutherford trajectories and we add the Coulomb field of the
partner.

{ The fluctuations of $\varepsilon (t)$ do not exceed 7\% and then
the possible excitation of vibrational modes is small and does not
affect the analysis.} Fig.~\ref{phi_r} shows that the evolutions
of $\varphi (D)$ for classical and TDHF calculations have the same
behavior. The maximum
reorientation predicted by this TDHF\ calculation is $\Delta \varphi =28.2{%
{}^{\circ }}$ ($33.6{{}^{\circ }}$), with the Sk$M^{*}$ (SLy4)
force. Both parametrizations give the same order of magnitude
($\sim 30{{}^{\circ }}$), however the classical expectation was
$\Delta \varphi \sim 23{{}^{\circ }}$. This $30\%-$difference,
which is independent on the initial orientation as we have
numerically checked, indicates a smaller moment of inertia in TDHF
as compared to the rigid-body classical approximation. This
reduced inertia can be attributed to a spherical core in the
N-body wave-function of the $^{24}$Mg which does not participate
to the rotation.

\begin{figure}[tbp]
\begin{center}
    \epsfig{figure=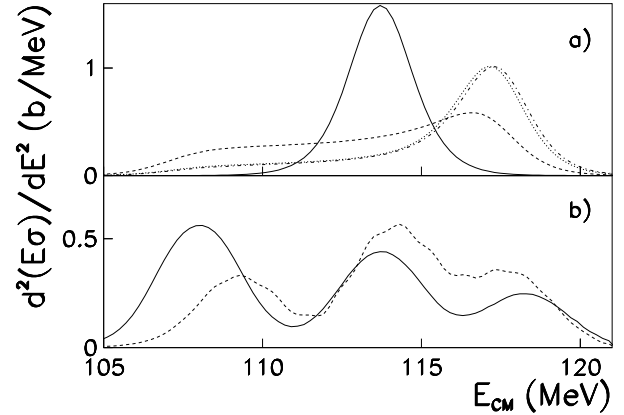,width=8.cm}
\end{center}
\caption{Barrier distribution for the reaction
$^{24}$Mg$+^{208}$Pb a) i- assuming spherical nuclei (solid line);
ii- considering a prolately deformed $^{24}$Mg ($\beta_2 = 0.4$)
and the AOF
 (dashed line); iii- and including reorientation
 with (dotted-dashed line) and without the rotational energy
(dotted line); b) CCFULL results without (solid line) and with
Coulomb coupling up to a distance of 241 fm (dashed line).}
\label{dist}
\end{figure}

Experimentally, the question is: what is the effect of the
reorientation on fusion cross-sections? { A first qualitative
investigation has been performed using} the code CCDEF
\cite{fer89} to estimate the fusion cross-section $\sigma
_{fus}(E)$ for the reaction $^{24}$Mg$+^{208}$Pb. CCDEF takes into
account the shape of the nucleus on the basis of the AOF. We then
go beyond this assumption by including in CCDEF the reorientation
obtained with TDHF. A commonly used way to present this excitation
function is to compute the so-called barrier distribution
$B(E)=\partial_{E^{2}}^{2}\left( \sigma _{fus}(E).E\right) $
\cite{row91}. Fig.~\ref{dist}-a shows barrier distributions
extracted from CCDEF without shape effect (i.e. the 1D barrier,
solid line). The width of the peak results from quantum tunneling.
A prolate deformation $\beta _{2}=0.4$ of $^{24}$Mg with an
isotropic distribution of orientation (dashed line) flattens
considerably the barrier distribution with a prominent part on the
high energy tail. A low energy shoulder extending down to $\sim$~5
MeV below the 1D barrier maximum is responsible for sub-barrier
fusion enhancement (as compared to the single barrier case).
Classically, the low energy part of the barrier distribution can
be interpreted as coming from ''parallel collisions'' whereas the
high energy part comes from ''perpendicular collisions''. The high
energy component dominates because a prolate nucleus has one
elongated direction (low barrier) for two short axis (high
barriers).

\smallskip The barrier distributions including the reorientation predicted
by TDHF  are plotted in Fig.~\ref{dist}-a both neglecting (dotted
line)  and taking into account the rotational energy { in the
trajectory} (dotted-dashed line) which is a second order
correction  in $\varepsilon$ and then increases only slightly the
barrier. Compared to the AOF prediction the TDHF results exhibit
reduced low energy shoulder and increased high energy peak. This
arises from that the Coulomb reorientation increases $\varphi$ and
thus increases the barrier height.

{ This phenomenon is implicitly taken into account in CC codes
like CCFULL \cite{hag99} when one uses big enough network to
include long range Coulomb couplings ($\gtrsim 200$ fm). Fig.
\ref{dist}-b shows CCFULL results
including nuclear (solid line) and nuclear+Coulomb (dashed line) couplings
to the five first excited states of the $^{24}$Mg rotational band with a
long range (241 fm). Marked variations appear in the barrier distribution
due to the fact that only the lowest part of the rotational band is
effectively excited \cite{row91} ($2^{+}-6^{+}$ in the present case). We
observe, like in Fig. \ref{dist}-a a decrease of the low-energy component
compensated by an  increase of the high energy part of the barrier
distribution. The previous TDHF+CCDEF results allow us to interpret these
variations as an effect of the Coulomb reorientation.}

Consequently, the sub-barrier fusion enhancement observed for reactions
involving a light nucleus on a deformed heavy ion like $^{154}$Sm is
expected to be reduced when the deformed nucleus is light and its collision
partner heavy. Experimentally, the effect of the reorientation should then
be studied by comparing the excitation functions of reactions with deformed
projectiles such as $^{24}$Mg on different targets. To simplify the
understanding of the reaction doubly-magic spherical targets such as $^{16}$%
O, $^{40}$Ca and $^{208}$Pb might be first tested. Eq. \ref{reorientation}
shows that the reorientation should increase with the mass of the target
reducing the sub-barrier fusion cross section.

Finally one may worry about the effect of the nuclear {
reorientation (implicitly included in CCFULL).} We have simulated
the fusion of two nuclei in TDHF. Since TDHF does not allow
directly tunneling, we have studied the modulation of the
threshold
energy $B$ of fusion reaction as a function of the initial orientation $%
\varphi $: $\Delta B(\varphi )=B(\varphi )-B(0)$. { The } deformed
nucleus { is } again $^{24}$Mg. To focus on the nuclear
contribution we { choose } a light spherical target, the $^{16}$O,
and we
{ start } the reaction at short distance. The observed variation of { B%
} appears to follow, within the numerical error due to the
considered energy step, the $\sin ^{2}\varphi $ modulation
expected for a quadrupole deformed
projectile in absence of reorientation. Indeed, we get $\Delta B(45{%
{}^{\circ }})/\Delta B(90{{}^{\circ }})=0.45(5)$ i.e. an almost
negligible deviation from the $\sin ^{2}$ law which predicts 1/2.
Inverting the problem to extract the reorientation leads to
$\Delta \varphi (45{{}^{\circ }})\!\sim \!-3{{}^{\circ }}$.
Considering the short range nature of the nuclear force this
effect is expected to not vary much with the target. Compared with
Coulomb effect this nuclear reorientation appears to be
negligible. This can be related to the range of the forces.
Indeed, the Coulomb interaction is a long range force so the
induced torque has time to rotate the nucleus and to produce a
large reorientation which is proportional to an average angular
velocity times the average time it is rotating. Conversely, the
nuclear field acts over a very short time so that even if it
contributes to the excitation of rotational states, the nucleus
has hardly enough time to actually rotate.

To summarize, we have studied the reorientation effect of a
deformed projectile on a spherical target. Analytical results for
the classical dynamics of a rigid body, confirmed by exact
simulations, show that, in contrast to a naive expectation, the
Coulomb reorientation depends neither on charges nor on relative
energy. The relevant observables are the deformation parameter and
the { inertias.}
 Those conclusions have been
extended to quantum dynamics using TDHF. These calculations show
that the nuclear contribution are negligible and they exhibit a
sizeable increase of the reorientation as compared to classical
calculations, interpreted in terms of a smaller moment of inertia
for the quantum system as compared to the rigid body
approximation. The reorientation is expected to be maximum when
the deformed nucleus is light and its collision partner heavy. {
For these systems, long range Coulomb couplings have to be
included in
  CC} calculations of barrier distributions { which } show that
sub-barrier fusion is partially hindered by the reorientation
process. We also suggest experiments to measure the effect of the
reorientation on excitation functions.

J.P. Wieleczko, N. Rowley, V. Yu Denisov and E. Pollacco are thanked for
fruitful discussions, M. Faure for his programming work and P. Bonche for
providing his TDHF code.

\end{document}